\documentclass[12pt]{article}
\usepackage[a4paper, margin=1in]{geometry}
\usepackage[T1]{fontenc}
\usepackage{lmodern}
\usepackage{setspace}
\usepackage{mathtools}
\usepackage{amsmath,amssymb,amsfonts}
\usepackage{bm}
\usepackage{graphicx}

\usepackage[round,authoryear]{natbib}
\setcitestyle{aysep={}}
\usepackage[dvipsnames]{xcolor}
\usepackage[british]{babel}
\usepackage{csquotes}
\usepackage[colorlinks=true,
    linkcolor=MidnightBlue,
    citecolor=MidnightBlue,
    urlcolor=MidnightBlue,
    filecolor=magenta,
    breaklinks=true
]{hyperref}
\usepackage{fancyhdr}
\renewcommand{\subsectionmark}[1]{}

\title{No Support Beyond Reach: The Measure Centrality Index 2.0 and Adversarial Comparisons of Consciousness Theories}
\author{Robert Chis-Ciure, PhD\\[2pt]
\small Sussex Centre for Consciousness Science \\
\small Department of Informatics, University of Sussex, United Kingdom}
\date{}

\begin{document}

\setlength{\abovedisplayskip}{7pt plus 1pt minus 1pt}
\setlength{\belowdisplayskip}{7pt plus 1pt minus 1pt}
\setlength{\abovedisplayshortskip}{4pt plus 1pt minus 1pt}
\setlength{\belowdisplayshortskip}{4pt plus 1pt minus 1pt}

\maketitle
\thispagestyle{empty}

\begin{abstract}
\noindent
A success in adversarial testing of scientific theories reaches only as deep as its matching failure would have cut. I call this the Reach Principle, and the goal of this chapter is to build the instrument that enforces it. Adversarial tests of consciousness theories can turn into a one-way evidential ratchet. When a preregistered prediction passes, the success is claimed for the research program; when it fails, the failure lands on an auxiliary involved in measurement or implementation. The ratchet turns wherever attribution is settled after the outcome is known. Against it, I define a preregisterable test object, putting the attribution policy on the record before the result, comprising what is compared, under which assumptions, when data suffice to settle an outcome, and which targets of a dated theory version each outcome exposes. The Measure Centrality Index 2.0 extends the original version to the positive-evidence side, recording for every possible outcome what it would count against and what its proponent has agreed to change. I run the framework on Cogitate's third prediction, where the preregistered target was the posterior-connectivity implementation and the mathematical or computational core of both Global Neuronal Workspace Theory and Integrated Information Theory stood outside the test. The record shows that a pass could have been claimed at any depth, because nothing fenced success the way the failure was bound. The next preregistration can set reach in advance, and support for a theory then reaches only what was exposed ex ante.

\vspace{1.2em}
\end{abstract}

\newpage
\markright{Contents}
\tableofcontents
\vspace{2em}

\section{The one-way evidential ratchet}\label{the-one-way-evidential-ratchet}

Adversarial tests of theories of consciousness (ToC) can acquire a characteristic \emph{asymmetry}. When a preregistered prediction passes, the result may be described as evidence for the whole research program, whether Global Neuronal Workspace Theory (GNWT) and Integrated Information Theory (IIT), the pair compared by \citet{cogitate2025}, or others currently tested \citep{corcoran-haun-et-al-2026}. However, when the prediction fails, the result may be absorbed by the auxiliary apparatus that made it testable, such as an instrument's coverage, how a region of interest is drawn, the frequency band analyzed, or the estimator chosen. Thus, success travels upward through the chain of auxiliaries, but failure stops at the first it meets. I call this the \emph{one-way evidential ratchet}. Whichever way the data fall, the public, field-facing standing of the research program turns in one direction only. The ratchet operates whenever the attribution of an outcome to a theoretical target is settled \emph{after} the outcome is known. At that point, the interpreter can choose where the result lands, and the incentive is to send success up to the program and failure down to auxiliaries.

The first experiment in the adversarial collaboration between GNWT and IIT is a good example of this attribution risk. \citet[p.~134]{cogitate2025} distinguished each theory's mathematical or computational core from its proposed neurobiological implementation, describing the former as difficult to test directly but the latter as testable with current methods. They also anticipated that theory proponents could respond to challenging data by revising the implementation and retaining the core; that is the downward half of the ratchet, foreseen before any data arrived. The authors then weighted each prediction by its centrality to the theories (p.~135) and closed by asking for a formal framework that integrates evidence and accounts for prediction centrality (p.~141).

\citet[pp.~3, 6]{chisCiure2024} proposed the Measure Centrality Index (MCI) to address problems of this kind. The index ranks empirical measures by the theory-relative consequences of disconfirming measure-based predictions, on a gradient from zero change to replacement of the framework. The present work, which introduces the MCI 2.0, extends the index to the positive-evidence side. Where first version settled how deep a failure cuts, the new one specifies how far a \emph{success} reaches, and it holds the two to the same level.

The chapter focuses on Cogitate's third preregistered prediction, which aimed at a target inside the proposed neurobiological implementation of each theory, in the spirit of the older dispute over whether the neural correlates of consciousness (NCC) sit in the front or in the back of the cortex \citep{odegaard2017,boly2017}. Whether the data warranted the verdict on that prediction remains in dispute. Moreover, IIT's formal apparatus stood outside the test altogether, a point addressed in detail below. Together, these three features leave one and the same result with several possible interpretations. A failure can be read as counting against the NCC implementation, or as showing only that the data were not good enough to count against anything. A success can be claimed for the implementation, or for the theory as a whole, whose core the test never touched, even for GNWT. Nothing in the data decides among these interpretations; the attribution policy decides, and in the first experiment that policy was settled \textit{after} the results were known. That is why attribution is the question at stake here.

When does a success in adversarial testing support a theoretical target? Consider the symmetry the evidential ratchet breaks. A success counts for a target only if the matching failure would have counted against that same target, under the same contrast and the same background assumptions, and only if the adverse outcome is one the data could have warranted. A success also counts for any target sitting above the failure's deepest reach, since whatever a failure reaches covers everything above it. And the whole arrangement must be in place before the result is known; the attribution policy that sends a success up has to be the same one that would have sent the failure down. 

Therefore, a success reaches only as deep as the matching failure would have cut; I call this the \emph{Reach Principle}. It constrains where support for a ToC may be claimed in public, but says nothing about how much, and leaves theory revision entirely free. That freedom invites an obvious evasion, namely revising the theory after an awkward result and claiming afresh. But revision does not erase the outcomes of adversarial tests. The preregistration lays down in advance everything that determines where an outcome lands, such as the contrast being tested, the background assumptions, what each possible outcome means, when the data suffice to call one, and which levels of the theory answer for the result. That bundle is the \emph{test object}. Change any part of it after the result is known, and you have made a new test object, one that meets future data on its own preregistered terms. The same holds on the theory side. Revising the theory's content, or the path it put on record from its commitments to the measurement, creates a successor theory-complex, which is obviously not identical to the initial version tested. By themselves, both moves are legitimate. However, what they cannot do is undo the record. The completed adversarial test reached what it reached, and a successor, whether a new test object of the same theory or updated theory-complex, begins with support only for the parts it carried over unchanged from the previous testing.

For example, an experiment may test a prediction about the NCC and leave the theory's core outside the failure's reach. Whatever result the data then warrant is an implementation-level result, however central the tested prediction sits in the theory's graph. An adversarial collaboration arranged this way can still run an informative experiment; but the support it yields, and the claims made from it, stay at that level. Cogitate's Prediction 3 shows the diagnostic value of these consequences, and it shows what goes missing when a preregistration sets pass--fail criteria without saying which target answers for which outcome. The next preregistration is where all of this becomes public \emph{before} the data arrive.

All of this stays continuous with the MCI 1.0. \citet[p.~15]{chisCiure2024} already treated relevance counterfactually, asking what disconfirmation of a measure-based prediction would require, and linked greater centrality to stronger evidence against a theory. They also allowed for revision, proposing an adaptive strategy under which classifications are updated after disconfirming experiments and the theory change that follows. What is updated here is the timing and the grain. The profile of consequences is now set out prospectively, outcome by outcome, in the preregistration itself, so the same document that says how deep a failure cuts also says how far a success reaches. And the profile attaches to a dated version of the ToC (e.g., IIT 4.0 \citep{albantakis2023} instead of 3.0 \citep{oizumi2014}), which is why revision after the result creates a successor, not a fresh claim on the old test.

What is new here is what gets regulated. The comparison tools the field already has weigh how much support an outcome gives; none of them says where a warranted success may land, and that omission is what the ratchet exploits. The MCI 2.0 sets it down in advance, and the contribution is timely, because adversarial collaborations are multiplying and every one of them will face the same attribution step. Section~\ref{what-the-measure-centrality-index-is-for} builds the working parts, namely, the measure-role that bears an index value, the test object that goes on record, and the profile listing, for each possible outcome, what it would count against and what its proponent has agreed to change. Section~\ref{reciprocity-in-confidence-and-in-public-attribution} gives the argument behind the Reach Principle, first as a lemma inside standard confirmation theory, showing that a success whose matching failure was sent to an auxiliary confirms nothing deeper than that auxiliary, then as a reciprocity requirement carrying the same constraint into public attribution. Section~\ref{why-structural-centrality-cannot-set-the-ceiling} explains why the position of a prediction in a theory's graph, however central, cannot settle how far support travels. Section~\ref{putting-reach-on-record-without-locking-theories-in-place} shows that keeping this record leaves revision entirely free, since a revised theory or an amended protocol simply starts a successor that inherits only what it kept. Section~\ref{prediction-3-and-the-reach-of-the-test} runs Cogitate's Prediction 3 through the framework as a worked case. Section~\ref{objections-neighbors-and-limits-of-mci-2.0} weighs the MCI 2.0 against its neighbors and states its limits, and Section~\ref{stopping-the-ratchet} closes with the rule itself: no support beyond reach.

\section{What the Measure Centrality Index is for}\label{what-the-measure-centrality-index-is-for}

Section~\ref{the-one-way-evidential-ratchet} ended with the promise that a preregistration can state in advance how far every possible outcome reaches. This section builds the machinery that would keep it. Thus, it specifies what a theory graph is and why centrality carries no meaning there on its own, what kind of thing bears an MCI value, what a test object collects, and what the MCI 2.0 profile assigns to each outcome.

A theory graph is a picture of a theory as a network. The nodes are the theory's claims, from its deepest commitments through the auxiliary hypotheses that connect them to experiments, down to the testable predictions, and an edge runs between two claims when one holds in virtue of another \citep[pp.~3, 8]{negroEtAl2026}. A centrality index scores the nodes of such a network by their structural importance. But every standard index encodes assumptions about what flows through the graph and which node outcome is being modeled; apply one to a process with different characteristics and it either loses its interpretation or gives poor answers. Node importance is always importance for an assumed process \citep[pp.~55--56,~69--70]{borgatti2005}. Thus, in a theory graph, what travels along the edges is evidence, and it travels only under an attribution policy. That policy is what gives the MCI its meaning, and so it has to be made explicit.

What bears an MCI value is a \emph{measure-role} ($\mu$). Let \(T^{v}\) be a specific, dated version of a ToC and let \[\mu=\langle m,\pi,T^{v}\rangle,\] where \(m\) is the empirical measure on focus and \(\pi\) is the path from the theory's commitments, through constructs, auxiliaries, and predictions, to an operationalization that \(m\) realizes. Every link in $\pi$ states what kind of dependence it asserts, whether a construct definition, an auxiliary hypothesis, or an operationalization choice, and every link points to where the theory or the preregistration says so. \citet[pp.~2, 5]{fazekas2024} separate a theory's central assumptions from the auxiliary hypotheses that connect them to implementation and measurement, and they observe that revising auxiliaries after a refuted prediction while keeping the core produces a new version of the entire theory-complex. The path \(\pi\) records exactly which of those links a given measure depends on. For instance, pairwise phase consistency (PPC) between category-selective and early visual sources is one measure. The role it plays for IIT's posterior-connectivity NCC implementation in the 2022 preregistration, where the estimator ran over frequency bands up to and including gamma, is one measure-role; the role the same estimator would play under a successor that moved the prediction to lower frequencies is another.

Evidence travels from data to theory only through an argument. A claim of empirical support ties together a phenomenon, an observation of it, and an interpretation, and directs them at a theoretical claim \citep[pp.~7--8]{kirkebyHinrup2024}. Evaluating such a claim means asking whether the argument is sound, whether it stays faithful to the empirical report, whether it can be tested, and how much of it runs on the theory's own vocabulary. All of that comes before the MCI and stays outside it. The index takes over only after the data warrant calling an outcome, and it says what follows from that call.

The prospective profile of post-testing consequences is also where two versions of a theory family part company. Consider versions A and B of one ToC family. They share everything an experiment can see: the theory graph, the prediction, the operationalization, the measure, the analysis plan, and the partition of possible outcomes. They differ only in where results land. Under A, a favorable result counts for the neurobiological implementation on the NCC side, and a matched adverse result counts against it; success and failure meet at the same level. Under B, the same favorable result is presented as support for the mathematical or conceptual core, while every matched adverse result is blamed on the estimator (e.g., the proxy for \(\Phi\) in IIT, or the decoder standing in for global ignition in GNWT). B is the evidential asymmetry mentioned above, adopted in advance. And the graph is identical across the two versions, every centrality identical, so nothing in it can tell them apart. The difference is public attribution, nothing else.

We can now write out in full the test object introduced earlier, denoted as $\mathcal{V}^{v,\tau}$. Every symbol in it is formalism for something already on the table: \[\mathcal{V}^{v,\tau}=\langle T^{v},\mu,\Pi_{\tau},B_{\tau},W,\Gamma,N,\mathcal R,\operatorname{Ctr},\preceq_{v},\rho,t_{v}\rangle.\] \(T^{v}\) and \(\mu\) are the dated theory version and the measure-role introduced above. \(\Pi_{\tau}=\{E_{1},\dots,E_{n}\}\) is the partition of possible results, finite and set in advance; each \(E_{i}\) is one way the experiment can come out, and I will call the members of the partition cells, so that every possible result falls into exactly one cell. A residual cell collects inconclusive data, failed measurements, and everything else the models allow, so that no result is left without a place to land; thus, \(\Pi_{\tau}\) is just an outcome space. Cogitate worked with a coarse partition of this kind when it sorted the possible patterns of results into PASS, FAIL, and CHALLENGE, the last one for results that tell against a theory but stay inconclusive given the limits of current methods \citep[pp.~42--43]{accelerating2022prereg}. \(B_{\tau}\) holds the background assumptions under which the comparison is made, the things both sides take as given rather than put under test; in Cogitate's case, these included the assumption that subjects consciously perceived every stimulus regardless of task demands, checked in a separate control experiment, not inside the test itself \citep[p.~22]{accelerating2022prereg}. \(W(e,E_{i})\) is the warrant rule. Data never settle an outcome on their own; someone must say in advance how strong the evidence has to be before the experiment counts as having come out one way rather than another. \(W\) is just that standard, written down per cell, saying when a body of evidence \(e\) suffices to conclude that \(E_{i}\) is what happened.

The next four elements state what follows once an outcome is in hand. \(K\) is the target put forward for support; \(H\) is the contrast, the alternative the target is compared against. \(\Gamma(E_{i};B_{\tau})\) is the support rule. Once \(W\) is satisfied and \(E_{i}\) stands, \(\Gamma\) says which target, against which contrast, may claim the success; in the A-and-B example above, version B directs the favorable cell at the mathematical or conceptual core, while version A at the NCC implementation. \(N(E_{i};H,B_{\tau})\) is the adverse rule, \(\Gamma\)'s negative counterpart, dictating which targets the same outcome counts against, relative to the same contrast and background. Under A, the matched adverse cell reaches the neurobiological implementation; under B, only the estimator. \(\mathcal R(E_{i})\) records what a proponent has agreed to change if \(E_{i}\) arrives, such as revise the theory, withdraw a commitment, relabel a link in the path, or restrict scope. The rule \(N\) and the record \(\mathcal R\) stay separate, because saying what an outcome counts against is weaker than agreeing to change something, and the probabilistic argument of the next section touches only the first.

The last four elements are what the Reach Principle runs on. \(\operatorname{Ctr}_{K,H,B_{\tau}}(E^{+})\) is the matching relation: for each favorable cell \(E^{+}\), it lists the adverse cells the protocol treats as that success's matching failures, meaning the results that would have counted as this very claim failing, under the same contrast and the same background. Without the pairing, ``the matching failure'' would be an empty phrase, and the principle could not be checked. \(\preceq_{v}\) is the depth order that makes ``deeper'' formally well-defined. It ranks targets by how far into the theory version's preregistered empirical identity a failure against them would reach, where the identity of a version is the bundle of commitments that make it this version rather than a successor. A failure that unseats only an estimator reaches shallow; one that would force the version to give up a commitment it cannot lose while staying itself reaches as deep as the order goes. Where two targets would reach equally far, the order treats them as one, and from here on target symbols stand for those classes. The order tracks identity reach alone; it ignores how far apart nodes sit in the theory graph, how much evidence an outcome carries, and how costly a revision would be. Finally, \(\rho\) and \(t_{v}\) make ``dated version'' precise. \(\rho\) is the provenance map, pointing, for every element above, to the passage of the theory or the preregistration where that element is stated, so a later dispute about what was agreed has somewhere definite to look. \(t_{v}\) is the timestamp, the moment after which a change to any element no longer edits the test object but starts a successor, which meets future data on its own preregistered terms.

MCI 2.0 is the profile that assigns, to each possible outcome, what that outcome would count against and what someone has agreed to change: \[\operatorname{MCI}_{v}(\mu,\tau;H,B_{\tau})=\{E_{i}\mapsto\langle N(E_{i};H,B_{\tau}),\mathcal R(E_{i})\rangle\}_{i=1}^{n}.\] Conceptually, the equation says that, if we take a measure-role \(\mu\) at preregistration time \(\tau\), a contrast \(H\), and the background \(B_{\tau}\), the index is then a table with one row per cell of the outcome partition, and the row for \(E_{i}\) carries two entries, namely what that outcome would count against (\(N\)) and what its proponent has agreed to change (\(\mathcal R\)). The MCI is nothing over and above that table of consequences. Nothing in it waits for data; the whole profile exists before the experiment runs. The four labels of the original version, ``orthogonal,'' ``peripheral,'' ``mantle,'' and ``core,'' become summaries of this profile wherever the targets in play form a clean nesting. To illustrate, a measure whose adverse cells reach only an auxiliary is peripheral; one whose adverse cells reach a commitment the version cannot lose and stay itself is core. Where the depth order leaves branches incomparable, the labels remain unassigned, and the profile itself is the answer. Unlike other tools \citep[e.g.,][]{corcoran2023, grim2022, henderson2010}, the MCI does not operate with likelihoods, Bayes factors, and posterior support, and it does not measure predictive success. Besides, how well a design discriminates between rivals is a property of the design and the models together \citep[ pp.~499--500]{myungPitt2009}, outside the index's scope. What the MCI records is where warranted outcomes land.

\section{Reciprocity in confidence and in public attribution}\label{reciprocity-in-confidence-and-in-public-attribution}

\subsection{The No-Free-Confirmation Lemma}\label{the-no-free-confirmation-lemma}

This subsection is the first of three steps toward the Reach Principle in its general form. It shows that the symmetry the principle demands is already enforced, in miniature, by probability theory, since inside a fixed statistical model, whatever can gain from an outcome can lose from another. The two subsections that follow carry that symmetry out of the model and into public attribution, and then generalize it to targets that depth cannot rank on a single ladder.

Within a fixed model, probability theory already enforces a version of the symmetry the Reach Principle demands. A model, here, is the complete pre-data setup of a comparison: background assumptions \(B\), a target \(K\), a contrast \(H\), and, for each of the two, a proper predictive distribution over the same space of possible outcomes, that is, a statement of how probable each outcome would be if that side were right. ``Fixed'' means the setup is chosen before the data and held constant while they are read; change the partition, the background, or either distribution, and you are running a different comparison. The other arrangements, a hierarchy of levels with agreed conditional probabilities or no probability model at all, come up later in this section. We suppose an outcome region \(E\) favors \(K\) over \(H\), which in this vocabulary says only that \(E\) is more probable if \(K\) is right than if \(H\) is: \[P(E\mid K,B)>P(E\mid H,B).\] Now consider \(E^{c}\), the complement of \(E\), which collects every other way the experiment could come out. Each side's probabilities across all possible outcomes must sum to one, so a side that makes \(E\) more probable must make the rest of the outcomes less probable overall. Writing that out, \[P(E^{c}\mid K,B)=1-P(E\mid K,B)<1-P(E\mid H,B)=P(E^{c}\mid H,B),\] the target that gains from \(E\) arriving is exactly the side that loses if \(E\) fails to arrive. The same holds at a finer grain. For a finite exhaustive partition \(\{E_{1},\dots,E_{n}\}\) of the kind \(\Pi_{\tau}\) collects, we write, for each cell, the gap between what the target expects of it and what the contrast expects of it: \[\delta_{i}=P(E_{i}\mid K,B)-P(E_{i}\mid H,B),\qquad\sum_{i=1}^{n}\delta_{i}=0.\] A cell with positive \(\delta_{i}\) is one the target treats as more probable than the contrast does, so observing it favors the target. The zero sum is the whole point. The \(\delta_{i}\) are differences between two collections of probabilities that each total one, so across the partition they cancel, and a positive \(\delta_{j}\) in one cell forces a negative \(\delta_{k}\) in another. In words, a model in which some outcome would support the target is, by that very fact, a model in which some other outcome would tell against it. If the two predictive distributions coincide, every \(\delta_{i}\) is zero, no cell favors either side, and the test is inert for this pair; a theory that expects exactly what its rival expects cannot be supported over that rival by any result.

Belief follows the same arithmetic, through standard Bayesian updating. Where cell likelihoods are positive and prior odds finite and positive, observing \(E_{i}\) moves posterior odds toward the target when \(\delta_{i}\) is positive and away from it when \(\delta_{i}\) is negative, with the usual convention when a likelihood is zero, under which the ratio counts as zero or as infinite depending on which side vanishes. So there is no free confirmation inside a model. Whatever stands to gain from one cell of the partition stands to lose from another, relative to the same contrast and the same background. This is the probabilistic neighborhood in which \citet{corcoran2023} place adversarial collaboration, where data arbitrate more strongly as the likelihood ratio moves away from unity (p.~3507) and precise predictions are, in their phrase, riskier bets that earn more evidence when they succeed (p.~3514).

The lemma only establishes existence, meaning that, somewhere in the same outcome space, under the same \(K\), \(H\), \(B\), partition, and pre-data model, there is a region that tells against the target. It does not say that anyone could recognize that region when it arrives. Moreover, the complement of a favorable cell may be scattered across the partition, with no single outcome a scientist could read as the counterresult. It also does not say the region is reachable in practice, since detecting it with available instruments, powering the study to reach it, and passing the warrant rule \(W\) all lie outside the mathematics. It sets no lower bound on how large the adverse \(\delta_{k}\) must be, it speaks only relative to the contrast inside the model, and it names neither where repair should land nor what anyone is obliged to revise. And where a theory has no predictive distributions at all, the lemma has nothing to grip; a verbal ToC goes unmodeled, which is an independent argument for mathematical consciousness science \citep{kleiner2020}.

\subsection{The Attribution-Reciprocity Principle}\label{the-attribution-reciprocity-principle}

The lemma of the previous subsection governs confidence inside a fixed model. Public attribution is a different matter, and it is where the evidential ratchet actually lives. This subsection states the norm that transfers the lemma's symmetry into public claims, and then defends it. In the Cogitate case, several candidate targets sit along the same path, and there is no model that assigns the outcome a likelihood under each target and rival. \citet{henderson2010} show what such a model would have to contain. In their hierarchical Bayesian picture, higher-level frameworks generate lower-level hypotheses through conditional probabilities \(p(T_{i}\mid T_{i+1})\) specified when the model is set up, and a framework is assessed only indirectly, through the empirical fit of the theories it makes probable (pp.~176--179, 198). Where such a hierarchy exists, the level an outcome reaches is computed. Where it is absent, the level is chosen. And a level chosen after the result is the evidential ratchet again.

Closing that gap requires a normative premise not to be found in the mathematics of ToC's adversarial testing. Consider the following:

\begin{quote}
If a public interpretation reads a warranted outcome as support for target \(K\) relative to contrast \(H\), the same pre-result attribution policy must identify at least one matched, warrant-eligible outcome of the same versioned measure-role that counts against \(K\), or against a target at least as identity-deep as \(K\), relative to the same \(H\) and \(B\).
\end{quote}

Why accept this premise? The defense rests on two invariances and one requirement of timing. The first invariance concerns the target. What a test is about does not change with how the result comes out. A policy that identifies the theory's core as what success supports has thereby named what a matched failure bears on; a test that can be read as a test of the core when it passes was a test of the core when it failed. The second invariance concerns the contrast. Favorable and adverse judgments have to compare the same alternatives under the same background assumptions. A success ``over the rival'' and a failure ``only for the estimator'' are judgments about two different tests. The timing requirement is by now familiar. The attribution policy must be settled \emph{before} the outcome is known. Choosing the level once the result is in is the move the principle exists to bar. Once everyone can see whether the prediction passed or failed, no choice of level is neutral again.

The No-Free-Confirmation lemma and the Attribution-Reciprocity principle do different jobs. The lemma lives inside a probability model and targets individual confidence, that is, to how strongly a researcher should believe a target given an outcome. The principle says nothing about belief. It concerns the public act that follows the experiment, when someone singles out which part of the theory the result supports. And even there the principle asks little. A failure may bear weakly on a core and heavily on an auxiliary, since magnitudes may be as asymmetric as the case warrants. A proponent may keep the supported target after any outcome. The whole requirement is one matched outcome, on record, that would have counted against the level claiming support. This means that before the result is known, the protocol must contain at least one cell such that, had the data landed there and passed the warrant rule, the proponent's own preregistered rules would have read it against the very target now said to be supported. Nothing more is asked: no promise to revise or abandon the theory, no symmetric magnitude, not even a probability model, only an address, specified in advance, at which the target could have been hit. A policy that cannot show such an address for a target never put that target at risk, and the principle then bars reading a success as reaching it.

\subsection{The Reach Principle under a partial order}\label{the-reach-principle-under-a-partial-order}

One step remains. The reciprocity principle speaks of targets at least as identity-deep as the one claiming support, and that phrase presupposes that targets can be compared for depth. But depth need not be a single ladder. One test may put in play a claim about the neurobiological implementation, a claim about a prediction's scope, and a constitutive claim about what the theory identifies consciousness with \citep{chisCiure2022transcendental, chisCiure2022dissertation, elliaChisCiure2022, chisCiureEllia2023hardProblem, chisCiure2024kant, chisCiure2025constitutive}, and any two of these may be incomparable under \(\preceq_{v}\). A workable rule has to handle such branches, so we build the ceiling in two steps. For a pair \(\langle K,H\rangle\) put forward for support, with \(K\) again the target and \(H\) the contrast it is compared against, and a favorable outcome \(E^{+}\), we first collect every target that the preregistered matching relation puts in the path of a matched failure, under the same contrast and background: \[X_{K,H,B_{\tau}}(E^{+})=\bigcup_{E^{-}\in\operatorname{Ctr}_{K,H,B_{\tau}}(E^{+})}N(E^{-};H,B_{\tau}).\] Conceptually, we take the observed success \(E^{+}\), look up every failure the protocol matched to it in advance, and pool everything those failures would have counted against. \(X\) is that pool, the full set of targets that were genuinely exposed while this success was still an open question. We then keep the deepest of them: \[\operatorname{Max}_{v}(X)=\{x\in X:\nexists\,y\in X\ \text{with}\ x\prec_{v}y\}.\] \(\operatorname{Max}_{v}\) goes through the pool and keeps a target only if no other member of the pool sits deeper than it. With both steps in place, the Reach Principle reads \[\langle K,H\rangle\in\Gamma(E^{+};B_{\tau})\;\Longrightarrow\;\exists\,x\in\operatorname{Max}_{v}\big(X_{K,H,B_{\tau}}(E^{+})\big)\ \text{such that}\ K\preceq_{v}x.\] That is, if the support rule lets the pair claim this success, then among the deepest exposed targets there must be at least one as identity-deep as \(K\). Support may be claimed at or below the deepest exposure, never above it. Sometimes the case yields a unique deepest exposed target, and the ceiling is that one target. Sometimes it yields several that the order cannot rank, and then the ceiling is all of them together; the supported target must sit at or below one of them. Averaging them, or electing one as the real deepest, is a category mistake. \citet[pp.~447, 460--461]{hand1996} proves the relevant point about ordinal structure, namely that an ordering licenses means or distances only with further structure. Hence, a preregistration for an adversarial test must say, for each of these deepest exposed targets, which outcome, contrast, and background expose it. And when the exposed set is empty, no target may claim support at all.

When the warrant rule calls an adverse outcome, that outcome counts against the preregistered version at the targets identified in \(N\), the adverse rule from the test object. A success reaches only as deep as its matched failures would have cut; the same contrast, the same background, the same measure-role. And because the profile of consequences is part of the test object, changing the profile changes the test, even when the theory keeps its name. None of this restricts what may be run; only what a run may be said to show. What remains is to say where the depth order comes from. The tempting answer is to read depth off the theory's own graph, from how central a node is; the next section argues that no such reading can set the ceiling.

\section{Why structural centrality cannot set the ceiling}\label{why-structural-centrality-cannot-set-the-ceiling}

If depth could be read off a theory's structure, the MCI 2.0 would be redundant, since a graph and a centrality algorithm would settle reach by computation. This section works through the closest thing the field has to that route, prediction maps, and locates exactly what a structural score can and cannot settle. Prediction maps show what the structural route looks like when carried through. \citet{negroEtAl2026} represent a theory as a directed graph whose nodes are source-linked claims and empirical predictions and whose edges are generic ``in virtue of'' dependencies, deductive or nondeductive in force (pp.~3, 8, 13). For every node, they compute local reaching, harmonic, betweenness, and PageRank centralities (pp.~9--12). In their IIT map, the node for content-specific posterior synchronization, Cogitate's Prediction 3, in focus here, has the highest PageRank of any neural prediction, 0.049, and they infer that it is the most important prediction to target in that map (pp.~19, 29). The authors cautiously note that the values depend on how the theory is reconstructed, that the maps omit some auxiliaries, that a map is one point in time, and that prediction and accommodation travel under one relation; theory challenge is what their method is after, with centrality one dimension of predictive informativeness among others (pp.~27, 30--31). None of those caveats matters here but the issue that does is of a different kind. Namely, a centrality score describes how a claim sits inside the theory's structure. It does not describe what a result may be taken to show. Thus, even if the map were perfect and every score exactly right, the scores would say only which predictions matter most to the theory. They would not say which part of the theory a passing result supports, or which part a failing result counts against. That question is answered by the attribution policy, and by nothing in the graph. The distinction matters because a high score invites precisely the slide the ratchet feeds on. The more central a prediction looks, the more natural it feels to interpret its success as support for the whole program, yet nothing in the score licenses that reading.

Recall versions A and B of a ToC, which share one graph and every centrality number, PageRank included. They differ only in where they send results. Under the Reach Principle, A's deepest exposed target is the implementation, so a success may be claimed there. For version B, \(X_{K,H,B_{\tau}}(E^{+})\) contains the estimator as its only deepest exposed target, and public support stops at the estimator or at measure-role performance. Could B do better? A fixed probabilistic model could place the favorable result and a reverse-likelihood region at core level, and the No-Free-Confirmation lemma guarantees that such a region exists whenever the favorable result favors the core over a rival. So the route to core-level support is open. To take it, the preregistration must place that reverse region in the outcome partition, match it to the theory-favorable cell, keep the same contrast and background, state the warrant standard, and put both the support and the exposure at core level. A preregistration that sends every adverse outcome to the estimator instead supports a different description, namely that the favorable outcome fits the full preregistered package, estimator included, and its public support stays at that level. The attribution policy thus sets the ceiling.

The point generalizes beyond a constructed pair. \citet{grim2022} model theories as Bayesian networks and show that the impact of evidence at a node depends jointly on the network's structure, the conditional probabilities on its links, the priors, and the likelihood character of the evidence. If one holds the first three fixed, changing the likelihood character alone can reverse which node has the greater network-wide impact (pp.~50, 53, 64). Now compare what each formalism knows. A prediction map records claim content, claim types, edge direction, and topology. Evidence sensitivity is determined by conditional probabilities, priors, and likelihood functions, which is exactly what the map leaves out. PageRank, for instance, recursively passes importance along incoming support links \citep[ pp.~11--12]{negroEtAl2026}; it sees the wiring, never the probabilities. So how far evidence reaches has to come from probabilistic structure where it exists, or from a separately preregistered attribution rule where it does not \citep{chisCiure2026inferential}.

The literature already interprets the selfsame prediction in two structurally different ways. \citet[p.~8]{negro2024} judges the three predictions of the first Cogitate experiment, Prediction 3 among them, to be quite peripheral, since they are formulated at the neural level and lean on many implementation-specific auxiliaries. \citet[p.~29]{negroEtAl2026} find that the very same node involved in Prediction 3 is the highest-PageRank neural prediction in their map. Both interpretations can be correct, because they measure different things, and neither settles where an outcome lands. The moral is to keep six quantities apart: structural centrality, evidential support, predictive success, discrimination, negative attribution, and revision obligation. The MCI 2.0 developed here records the last two, prospectively. The first four belong to other tools (more in Section~\ref{objections-neighbors-and-limits-of-mci-2.0}). Attribution policy, then, is not something a graph can compute; it has to be made explicit. The next section says where and how, and answers the worry that writing it down locks a theory in place.

\section{Putting reach on record without locking theories in place}\label{putting-reach-on-record-without-locking-theories-in-place}

The previous section ended with an obligation, that attribution policy be made explicit somewhere. This section says where, namely in the preregistration itself, and in what form; it then shows that keeping such a record takes away none of a theory's freedom to develop, because revision creates successors rather than erasing what was tested. Everything turns on what is settled in advance. The Cogitate Consortium preregistered its predictions, pass-or-fail criteria, expected outcomes, and their interpretation before the held-out data were analyzed \citetext{\citealp[ p.~133]{cogitate2025}; \citealp[ pp.~10--12]{accelerating2022prereg}}. That goes beyond a bare hypothesis-and-analysis preregistration, but it still stops short of what the ceiling needs. A generic preregistration pins down hypotheses, analyses, and outcomes. The Reach Principle asks for more: the background, the matched target--contrast pairs on the support side, and the negative attribution on the failure side. Concretely, the contrast-indexed profile \(\{E_{i}\mapsto\langle N(E_{i};H,B_{\tau}),\mathcal R(E_{i})\rangle\}\), with the support rule \(\Gamma\) and the matching relation \(\operatorname{Ctr}\) alongside it, has to sit inside the preregistered document. With those fields in place, every outcome cell arrives with its target-level consequences already on record.

One profile binds one completed test; theory development then proceeds through successors. A post-result change to the contrast, background, outcome meanings, warrant rule, matching relation, or attribution assignments creates a successor of the test object \(\mathcal V\). A change to the empirical path, intended scope, target content, or the theory's empirical commitments may create a successor theory-complex \(T^{v+1}\). Renaming and rearranging that preserve every outcome consequence create nothing new; the preregistered objects stay in place.

None of this short-circuits theory-building. A theory may repair an estimator, change a frequency hypothesis, alter a timing claim, relocate a substrate, restrict a scope, or revise the connector between its mathematical claims and their biological reading \citep{chisCiureLevin2025, zheng2026loops}. However, direct support passes to a successor only through what stayed unchanged, be that the supported target, the relevant path in full, the contrast, the background, the reading of outcomes, the warrant standard, and the attribution policy. Ordinary evidential appraisal then decides what transfers, and a changed component can claim direct support again once its altered obligation has itself been tested. The predecessor's warranted adverse result stays on the record of the object that ran it. \citet[pp.~294--295, 300]{cronbachMeehl1955} drew an analogous line for construct validation. When a prediction fails, any link in the inferential chain may be at fault, and altering the network to cope with the observations redefines the construct, which must then be validated on fresh data before being advanced publicly, on pain of substituting rationalization for validation.

The adaptive strategy of MCI 1.0  allowed a mantle measure that met disconfirming experiments, followed by theory change, to be downgraded to peripheral or dropped once its relevance lapsed \citep[ p.~15]{chisCiure2024}. The downgrade is legitimate, and it belongs to the successor. The predecessor keeps its measure-role inside its completed test, because the measure-role that was tested is the one the predecessor put on record. Similarly, \citet[pp.~33--35, 47--49]{lakatos1978} attaches appraisal to a series of theories, treats the protection of a hard core as a methodological decision of its proponents, and lets the protective belt of auxiliaries take the hits. He also grants that experiments are recognized as decisive only with hindsight, once a rival explains the earlier program's success (pp.~69--72). Still, what his series appraisal never gives is a verdict for the single moment, a rule saying what this success may claim now and what this failure counts against now. The successor rule introduced here does that.

Where do the adverse rule \(N\) and the revision record \(\mathcal R\) come from? One way is for proponents to commit themselves in advance. Before results are known, they state which responses to each adverse cell they renounce and which targets they accept as exposed. A structured elicitation such as the IDEA protocol organizes this well, with private initial judgments, feedback and discussion, private revision, and mathematical aggregation that keeps the uncertainty bounds \citep[ pp.~3, 5--6]{hemming2018}. The evidence for the protocol's accuracy concerns questions whose answers were later realized, but exposure assignments are answers of a different kind; the study also lacked a feedback-free control condition (p.~27). But commitment only builds the map. A theory's source texts still settle its constitutive core and the scope of its claims, independent assessment still settles logical consequence, source fidelity, and warrant, and exposure at the level of repairs needs evidence beyond a simple pass-or-fail preregistration. Hence, reconstruction from the theory's sources remains necessary.

Some cases want more, and something stronger is available from what we introduced above. Let \(A^{-}_{v}(E_{i})\) be the preregistered family of responses to a warranted adverse outcome, and let \(Pres_{K}(\rho)\) hold when response \(\rho\) preserves the target \(K\) and its preregistered scope. Revision closure ($RC$) is then \[\begin{aligned} RC(E_{i};K)=1\iff {}& A^{-}_{v}(E_{i})\neq\varnothing,\\ & A^{-}_{v}(E_{i})\ \text{is prospectively fixed and adequately bounded, and}\\ & \forall\rho\in A^{-}_{v}(E_{i}),\ \neg Pres_{K}(\rho). \end{aligned}\] The first condition asks that the preregistration list at least one response to the adverse outcome; the second, that the list be set before the data and adequately bounded, meaning it says why each member belongs to it; the third, that every listed response gives up \(K\) or its preregistered scope. Together they enforce that a proponent who faces this outcome has nowhere agreed to stand that keeps the target intact. \(RC=1\) means the test was decisive for \(K\), since every response left open changes \(K\) or its scope. \(RC=0\) means a target-preserving response stayed open, and the outcome may still have carried evidence against \(K\); preserving a target and having evidence against it can coexist. Ordinary theory testing proceeds without closure, and the Reach Principle is silent about that. The point of the distinction is to keep the MCI framework's demands modest. A test can be worth running, and its outcomes worth recording, without being decisive for any deep target. Closure marks the special case where proponents have agreed in advance that an outcome would leave the target no room, and everything else the chapter asks for works without it. The record-keeping is now fully specified: what goes into the preregistration, how successors inherit, and when a test is decisive. What remains is to run it on a real case, and that is Cogitate's Prediction 3.

\section{Prediction 3 and the reach of the test}\label{prediction-3-and-the-reach-of-the-test}

The preceding sections built the framework in general terms; this one brings it back to the case that motivated it, Cogitate's third prediction. The MCI 2.0 machinery introduced here is a diagnostic tool saying what the attempted test did and did not reach. Nothing is applied retroactively. To make the diagnosis, one must distinguish, for each theory, the target that was preregistered from the core that was not. \(T_{\mathrm{reg}}\) is the agreed NCC implementation. For IIT, it connects the theory's cause--effect account to sustained, content-specific synchronization between activated posterior cortical areas, under the PPC measure-role; for GNWT, it connects the workspace account to content-specific synchronization between prefrontal and category-selective areas in the 300--500 ms window in which the theory locates global broadcast, under the same estimator \citetext{\citealp[ pp.~6--8]{melloni2023}; \citealp[ p.~28]{accelerating2022prereg}}. \(T_{\mathrm{core}}\) is what stands behind each implementation and was not itself on the line: for IIT, the mathematical formalism, \(\Phi\) and the \(\Phi\)-structure \citep{oizumi2014, albantakis2023}; for GNWT, the computational architecture on which a content becomes conscious when the workspace broadcasts it globally. The consortium itself drew this line, describing the core as difficult to test directly and the neurobiological implementation as what current methods reach, and noting that an implementation can be revised while the core stands \citep[ p.~134]{cogitate2025}. Dehaene, writing for GNWT, placed \(\Phi\) and the rest of IIT's mathematical apparatus outside the experiment's reach \citep[ Supplementary Information, p.~123]{cogitate2025}. About GNWT's own core, he entered no such reservation, reading the transient prefrontal ignition, and the exploratory connectivity pattern accompanying it, as confirming the workspace account; moreover, he treated the adverse results as artifacts of electrode coverage and signal grain, and questioned the assumption this chapter used in Section~\ref{what-the-measure-centrality-index-is-for} to illustrate \(B_{\tau}\), agreed by all parties in advance, that subjects experienced each stimulus for its full duration \citep[ Supplementary Information, pp.~122--123]{cogitate2025}. The asymmetry between the two halves of that commentary, core bracketed for the rival, core confirmed for one's own theory, is the ratchet performed within a single record document. Furthermore, IIT's critics argue, for different reasons, that common empirical proxies measure quantities distinct from the theory's mathematical quantity, which leaves the core claims unestablished \citep[ pp.~689--691]{iitConcerned2025}. Thus, all parties place IIT's \(T_{\mathrm{core}}\) outside Cogitate's adversarial collaboration. The dissensus is about what follows from that.

The diagnosis below returns to the dissensus and gives both interpretations their due. First, the record itself needs establishing, and one feature of it is relevant later, namely the registration history. The protocol by \citet[pp.~1, 6--8]{melloni2023} recorded predictions endorsed by the theories' proponents, and it deliberately left the analysis specification open until an optimization phase had run on a first portion of the data; the analysis would then be locked, and the remaining data used as the test. Version 4.5 of the preregistration, dated 22 December 2022, is that locked specification. It was amended after analyses were developed on the first third of the Experiment 1 data, and the remaining two thirds of the magnetoencephalography (MEG) and functional MRI (fMRI) data were held out under it. The one-third/two-thirds split applied to MEG and fMRI alone; intracranial EEG (iEEG) followed a different strategy, forced by limited and unpredictable patient availability \citetext{\citealp[ pp.~1, 10--12]{accelerating2022prereg}; \citealp[ Methods, Participants]{cogitate2025}}. So the final criteria were \emph{prospective} for the held-out MEG and fMRI analyses, while some dataset and analysis choices were made \emph{after} data inspection had begun. The claims below take that explicitly into account.

Prediction 3, as preregistered for IIT, required stronger sustained synchronization, from 300 ms after stimulus onset until its offset, between activated V1/V2 and the fusiform face area (FFA) for faces, or between activated V1/V2 and the lateral occipital complex (LOC) for objects, for all stimulus durations, in MEG or iEEG, with the difference in synchronization pattern more consistent with the stimulus than with the task \citep[ integration table, p.~28]{accelerating2022prereg}. On duration, the preregistration disagrees with itself. Specifically, the integration table requires the effect across all durations, but the detailed iEEG and MEG PPC plans exclude the 0.5-second trials. Whether the implemented PPC test adequately realizes the obligation as the integration table states it is itself a question for the warrant rule \(W\). Every ambiguity of this kind widens the room for after-the-fact interpretation, and the dissensus documented below lives exactly there.

PPC between category-selective channels or sources and V1/V2 operationalized the phase relation, computed across low and high frequency ranges, with iEEG channels included only if they showed sustained increases in high-gamma activity across three preregistered time windows \citep[ pp.~61--62, 67--68]{accelerating2022prereg}. The frequency band is where the measure-role as preregistered and its later interpretations diverge. The integration table left gamma outside the pass-or-fail criteria, yet the published article describes IIT as predicting sustained gamma-band connectivity, explains that gamma was analyzed for its link to spiking, and reports that the observed activity was transient and outside the gamma band \citep[ p.~139]{cogitate2025}. The theory proponents describe their prediction as concerning synchrony in the gamma range, and then suggest that the prediction may be wrong about the frequency range (Supplementary Information, p.~121). A path \(\pi\), encoded in the MCI 2.0, with labeled links would have recorded whether the frequency claim belonged to the preregistered obligation or to commentary laid over it. Its absence is one reason the case is instructive for developing the MCI as a standalone tool.

All parties accept the preregistered outcome. Under the PPC analysis, both IIT and GNWT failed Prediction 3. Posterior effects were early and brief, restricted to low frequencies, and largely explained by the stimulus-evoked response \citep[ p.~139]{cogitate2025}. The consortium, from the position of a theory-neutral party, called the absence of sustained posterior synchronization IIT's ``most direct challenge'' under the preregistration, and it acknowledged that the null result could stem from limited iEEG sampling of V1/V2 \citep[ p.~140]{cogitate2025}.

As IIT's proponents, Boly, Koch, and Tononi accepted the lack of support for Prediction 3 but disputed whether the data were adequate to show it. According to them, twelve electrodes sat in V1/V2 against several hundred in prefrontal cortex, so the areas the prediction actually concerns were barely sampled, and the channels that were there had been let in by an eligibility rule loosened after preregistration to include ones without sustained activation. The MEG regions fare no better in their view, since they were built from broad-band evoked fields, which mix activated with deactivated tissue, whereas the preregistered prediction concerns activated areas only. And the synchrony itself, they suggest, may exist at lower frequencies than the analysis emphasized \citep[ Supplementary Information, pp.~119--122]{cogitate2025}. The consortium's own record of deviations lists the relaxed eligibility rule as a departure from the preregistration, alongside changes to fMRI pooling and multiplicity correction, and it notes that dynamic functional connectivity (DFC) was added only after the phase-based analysis found connectivity absent \citep[ Supplementary Information, pp.~124--125]{cogitate2025}, with DFC analyses being exploratory. In iEEG on task-irrelevant trials, the consortium interpreted them as consistent with GNWT and as challenging IIT \citep[ p.~139]{cogitate2025}; yet its summary of the task-relevant analyses found them inconsistent across modalities and short of a conclusive argument for either theory's prediction \citep[ Supplementary Information, pp.~95--96]{cogitate2025}.

This is exactly the situation MCI 2.0 exists for. Recall what the index attaches to a measure-role, i.e., the consequence profile, the outcome-by-outcome table from Section~\ref{what-the-measure-centrality-index-is-for} saying in advance what each result would count against and what would have to change. That is what the attempted test lacked. Cogitate's preregistered materials cover the empirical prediction and its pass--fail criteria, but the Reach Principle asks for four further fields: the background \(B_{\tau}\) both sides take as given, the support rule \(\Gamma\) saying who may claim a success, the adverse rule \(N\) saying what a failure counts against, and the matching relation \(\operatorname{Ctr}\) pairing each success with the failures that answer it. Pieces of \(B_{\tau}\) do appear, scattered through the materials as assumptions and control checks, the conscious-perception assumption of Section~\ref{what-the-measure-centrality-index-is-for} among them; the rest never appears: no support rule, no adverse rule, no matching relation, and no assembly of the four into one profile \citep[cf.\ integration table, p.~28]{accelerating2022prereg}. The materials reviewed above settle the intended target of the attempted test, \(T_{\mathrm{reg}}\), and place \(T_{\mathrm{core}}\) outside it. Everything the diagnosis says is conditional on the warrant rule \(W\), the standard for when evidence suffices to settle an outcome, and its assessment lies outside the MCI. Here the dissensus lands. If the consortium's interpretation of the evidence stands, the public negative attribution reaches \(T_{\mathrm{reg}}\), together with the empirical connector inside it, the auxiliary claim that causal binding among the activated posterior areas shows up as sustained phase synchronization; \(T_{\mathrm{core}}\) stays outside, with no likelihood and no adverse assignment. If IIT's proponents are right, the evidence fell short of testing the intended NCC implementation, and nothing was reached at all. Notice what the two conclusions share. The intended target is the same under either judgment, so the whole disagreement sits inside \(W\), over whether the data warranted the outcome, and never spreads into where a warranted outcome would land. The test object's separation of warrant from attribution is what keeps it confined there. And any change to the eligibility rule, sampling density, region construction, estimator, or frequency claim produces a successor test object wherever the preregistered obligation changes.

Now the successor rule from Section~\ref{putting-reach-on-record-without-locking-theories-in-place} says that revision stays free, that a revised account meets future data as a new object, and that support carries over only through what stayed unchanged after testing. To wit, a lower-frequency account of causal binding, a DFC-based account, or an account tied to denser V1/V2 sampling is a successor wherever it changes the preregistered obligation. It may well be progressive, and it becomes eligible for direct support once the changed part is itself tested under a fixed comparison and a matched attribution map. The exploratory DFC fit belongs to the existing data and keeps its exploratory status. The same rule binds the other theory. The iEEG DFC pattern the article describes as consistent with GNWT's predictions stays exploratory for it, and GNWT's own preregistered Prediction 3 failed under PPC as well \citep[ p.~139]{cogitate2025}. \citet[p.~6]{fazekas2024} interpreted the proponents' responses as deducing the failure of the prediction from the failure of measurement-specific auxiliaries while defending a central claim. On the present MCI account, that response stays open, and it comes with a version boundary.

Could the ceiling for Prediction 3, the deepest target a success would have been entitled to reach, be computed now, from the record as it stands? Almost, and the near-miss is interesting. An exact ceiling for the success that did not happen, a statement of how deep a passing Prediction 3 could have reached, needs the full test object. Cogitate's preregistration identifies candidate implementation-level targets, the posterior substrate, sustained connectivity and its timing, and the PPC measure-role, so the raw material is there. What is missing is the contrast-indexed adverse rule \(N\) and the local depth order \(\preceq_{v}\), and neither can be reconstructed after the fact without deciding disputed questions by fiat. Frequency shows why. The final integration table left gamma unrestricted, saying nothing about which band the obligation ranged over, while the surrounding commentary treated gamma as the prediction's content; whether a lower-frequency success would have counted is exactly what a depth order would have had to settle, and no document settles it. The materials place \(T_{\mathrm{core}}\) outside the intended empirical targets, and the highest PageRank of the corresponding node in the local IIT prediction map records structural importance, nothing more \citep[ p.~29]{negroEtAl2026}. Therefore, the next preregistration is the first place where the deepest exposed targets can actually be computed.

What, in the end, may be publicly said about Prediction 3? Less than the field has been saying. The preregistered target was the posterior-connectivity implementation; if the evidence holds up, the result counts against that implementation, and nothing deeper. The more interesting finding, though, concerns the experiment that did not happen. Had Prediction 3 passed, nothing on record would have confined the success the way the failure is confined now: no field says where a success may land, and no matching relation pairs it with the failures that would have had to be risked. A pass would have been claimed as deep as ambition allowed, and the record could not have objected. That asymmetry, failure fenced in, success unfenced, is the ratchet's signature, sitting intact in the field's most prominent adversarial test. The gap is nobody's oversight, since no instrument existed for recording reach. That is MCI 2.0's raison d'etre.

\section{Objections, neighbors, and limits of MCI 2.0}\label{objections-neighbors-and-limits-of-mci-2.0}

The framework is now complete and exemplified. Next, we consider the strongest objections, place it among its neighbors, and specify its own conditions of validity. Hierarchical Bayesian modeling brings the strongest objection to the Reach Principle. A theory's core can make a successful implementation more expected, while on failure rational updating lands mostly on a fragile auxiliary. In a hierarchy of the kind \citet[pp.~176--179, 198]{henderson2010} describe, where a framework is judged through the hypotheses it generates, the core's adverse Bayes factor can be tiny while the auxiliary's is large. \citet[p.~3514]{corcoran2023} add that a theorist who puts a surprising result down to mitigating factors, and so changes their mind as little as possible, is behaving rationally by Bayesian lights. On this view, requiring core revision would reward brittle theories and misdescribe hierarchical updating.

The asymmetry is real, and nothing in the MCI 2.0 denies it. The Reach Principle allows a stable core, unequal updates, and unequal likelihood ratios. What it forbids is narrower than the objection assumes. If the core is to be supported in public when the experiment succeeds, then somewhere in the model's own outcome space there must be a region that would have counted against the core, because the No-Free-Confirmation lemma puts one there whenever the core genuinely raises the probability of the favorable result. The exposure may be as small as the hierarchy makes it; the principle cares that it exists and is on record, not that it is large. So a proponent with a genuine hierarchical model is nearly home already. The model meets the probabilistic condition by itself. What public support at core level still needs is the paperwork, namely the adverse region written into the outcome partition, matched to the favorable cells by \(\operatorname{Ctr}\), with the warrant standard \(W\) and the support and adverse rules \(\Gamma\) and \(N\) set out in advance. Where such a model exists and is on record, it is an independent road to theory-level support, and the principle raises no obstacle. The objection is therefore an argument for building the model. Until it is built and put on record, public claims answer to the Reach Principle, the rule that a success reaches only as deep as its matched failure would have cut.

A second objection to the Reach Principle says that preregistered exposure leaves no room for theories to grow. Theory change is normal; on Lakatos's account, the belt exists to be adjusted, and \citet[pp.~4, 8]{negro2024} expects proponents of IIT and GNWT to absorb contradictory evidence by revising auxiliaries and leaving the core intact, with both programs staying progressive as long as the revisions generate new testable content. The successor rule, that a revised theory or test meets future data as a new object and inherits support only through what stayed unchanged, leaves all of that untouched. What it holds still is the identity and the consequences of one completed test. Thus, a successor that outperforms its predecessor gathers support under the preregistration filed for it, and a series that keeps producing tested successors is progressive in exactly the sense of \citet[pp.~33--35]{lakatos1978}.

Still, commitments on behalf of ToCs can be gamed. A trivially low-level target produces a low ceiling, but the lowness is visible in the preregistration, and attribution is confined to that level for everyone to see. A nominal core-level exposure raises a different worry, about magnitude. Level and magnitude answer different questions. The level identifies the exposed theory-internal target; the magnitude measures the force of the evidence at it. The Reach Principle sets the level and leaves magnitude to the probabilistic model or to another account of confirmation, so a vanishing core-level exposure and a substantial one do meet the same level condition, and a gamed commitment licenses its target level and nothing else. Two checks remain. Reconstruction from the theory's source texts catches commitments that misdescribe the theory, and declining deeper exposure is itself informative, since it puts the experiment's scope on record.

A few remarks on the division of labor with similar tools in the field. The original MCI gives disconfirmation semantics and qualitative levels, while this 2.0 update adds a level constraint on positive support and explicit version inheritance. Structural relevance and prediction maps locate claims within a graph of the theory, the network of claims and dependencies from Section~\ref{what-the-measure-centrality-index-is-for}, and leave public outcome-to-target attribution to a separate map \citetext{\citealp[ pp.~6--7]{negro2024}; \citealp[ p.~27]{negroEtAl2026}}. Bayesian model comparison quantifies evidential magnitude where likelihoods exist \citep[ pp.~3509--3510]{corcoran2023}; the reach ceiling defended here constrains target attribution where a fixed cross-level model is absent. A generic preregistration pins down hypotheses, analyses, and outcomes; matched claim-level support and exposure are a further requirement, and the MCI 2.0 profile is the form that requirement takes, the extra fields a preregistration must carry before its outcomes can be assigned a reach. The classic logic of theory change, known as AGM after \citet[p.~510]{alchourron1985}, sorts the candidate ways of giving up a proposition while keeping as much of a theory as possible, through intersections of maximal non-implying subsets; whether success may be claimed above every preregistered failure target is a question it never asks. That question is within the remit of the Reach Principle. Thus, AGM says how to retreat once something must go; the Reach Principle says what a completed test entitles anyone to claim while everything still stands. And Lakatosian appraisal allows core protection and judges programs across versions, over time; the proposal here adds the missing rule for the single test, the Reach Principle, barring outcome-conditioned level switching.

The chapter's central result, the Reach Principle with the No-Free-Confirmation lemma behind it, holds under four conditions. First, the comparison must stay stable, meaning the same target, contrast, background assumptions, outcome partition, and predictive distributions across theory-favorable and theory-adverse cells. The probabilistic half of the argument works only inside such a stable comparison; where stability fails, the reciprocity norm of Section~\ref{reciprocity-in-confidence-and-in-public-attribution}, the requirement that support claimed at a level be matched by exposure at that level, carries the weight alone. Second, public outcome-to-target assignments must matter, in the sense that they change at least one warranted judgment, about evidential support for a theory, about the scope of the test, or about which test was run. Where a community's practice would be identical under any assignment, the principle constrains nothing, because there is nothing to constrain. Third, a finite local depth order must be reconstructable for the test at hand. The ceiling of a given success is a location in that order, so where no order can be built, there is no ceiling to compute, only the bare reciprocity requirement again. Fourth, and least obviously, settling attribution in advance must add something beyond what an adequate probabilistic model already settles. This is the redundancy condition, and it cuts in the framework's favor. A complete hierarchical model of the kind the Bayesian objection invokes would compute every level assignment on its own, and inside such a model the principle would have no work left; it would be true and inert. The principle matters exactly where that model is missing, which is to say in the actual practice of consciousness science, where no agreed cross-level model exists and level assignments are made in prose, after the results. Cogitate sits squarely inside that domain, and its record shows it, inasmuch as the consortium and theory proponents issued divergent public verdicts over one and the same outcome \citep[ p.~140]{cogitate2025}. Whether the parties can agree on a stable local order is a test the successor preregistration will run. The next section collects what all of this amounts to.

\section{Stopping the ratchet}\label{stopping-the-ratchet}

Wrapping up, for the successor test of Prediction 3 that the consortium urges IIT's proponents to pursue \citep[ pp.~140--141]{cogitate2025}, the design consequence is immediate. Its preregistration must state the comparison background, the contrast, the full outcome partition, the matching relation, the warrant standard, and the negative targets from which each deepest exposed target follows. Concretely, it must say, before the data arrive, whether a lower frequency band, a denser V1/V2 sample, or an amplitude-based estimator belongs to the successor of \(T_{\mathrm{reg}}\) or to \textit{post hoc} commentary that carries no exposure at all. If agreement stops at the posterior-connectivity NCC implementation, the report is set at that level in both directions; the experiment stays worth running, and whatever it confirms or challenges is the implementation. If the parties agree to expose a claim of \(T_{\mathrm{core}}\), say a claim about which physical properties of a substrate the theory holds constitutive of experience, that claim enters the adverse rule \(N\) for a given matched cell, and a pass, given the warrant standard \(W\) and the fixed contrast and background, may then support the theory at that level.

This settles a question the first experiment left open. A successor that changes the frequency claim and then passes gains direct support for the changed connector only if the result stands under the preregistered contrast, background, support assignment, and matched exposure. The predecessor keeps its failure on the record, and \(T_{\mathrm{core}}\) stays untested until a preregistration exposes it. Simply enough, proponents who want their theory's core supported by the next experiment must put the core within the next test's reach. Proponents who decline may keep the core out of reach, and a core kept out of reach stays outside experimental support.

What MCI 2.0 does is set these options side by side and put each on the record before the data arrive, the record being nothing grander than the preregistered profile itself, the document that says who may claim what. The chapter began with the one-way evidential ratchet, the asymmetry by which success is sent up to the research program while failure is absorbed by the auxiliaries. Against that asymmetry, it built a principle, an object, and an index. The Reach Principle says that a success reaches only as deep as its matching failure would have cut. The test object is the preregistered bundle that makes the principle checkable, saying in advance what is compared, under which assumptions, and where each possible outcome lands. MCI 2.0 is the profile that makes those consequences explicit, outcome by outcome, while every one of them is still a possibility. The timing is the whole point. The evidential ratchet turns because attribution is settled after the outcome; settle it before, and there is nothing left for it to turn on. Success then travels exactly as far as the matching failure would have, no farther and no less, along a path the theory itself agreed to. A core that wants support from experiments must come within their reach. Everything else is commentary, and the record will say so.

\begingroup
\hbadness=10000
\bibliography{refs}

@misc{accelerating2022prereg,
  key          = {Accelerating Research on Consciousness},
  title        = {Accelerating research on consciousness: An adversarial collaboration to test contradictory predictions of global neuronal workspace and integrated information theory},
  year         = {2022},
  month        = dec,
  howpublished = {OSF},
  note         = {Version 4.5, preregistration, 22 December 2022},
  doi          = {10.17605/OSF.IO/92TBG},
}

@article{alchourron1985,
  author  = {Alchourr{\'o}n, C. E. and G{\"a}rdenfors, P. and Makinson, D.},
  title   = {On the logic of theory change: Partial meet contraction and revision functions},
  journal = {The Journal of Symbolic Logic},
  year    = {1985},
  volume  = {50},
  number  = {2},
  pages   = {510--530},
  doi     = {10.2307/2274239},
}

@article{borgatti2005,
  author  = {Borgatti, S. P.},
  title   = {Centrality and network flow},
  journal = {Social Networks},
  year    = {2005},
  volume  = {27},
  pages   = {55--71},
  doi     = {10.1016/j.socnet.2004.11.008},
}

@article{chisCiure2024,
  author  = {Chis-Ciure, R. and Melloni, L. and Northoff, G.},
  title   = {A {Measure Centrality Index} for systematic empirical comparison of consciousness theories},
  journal = {Neuroscience and Biobehavioral Reviews},
  year    = {2024},
  volume  = {161},
  pages   = {105670},
  doi     = {10.1016/j.neubiorev.2024.105670},
}

@article{chisCiure2024kant,
  author  = {Chis-Ciure, R.},
  title   = {{Kant's} a priori: Between necessity and constitutivity},
  journal = {Philosophica: International Journal for the History of Philosophy},
  year    = {2024},
  volume  = {32},
  number  = {1},
  pages   = {19--40},
  doi     = {10.5840/philosophica202512220},
}

@article{chisCiure2025constitutive,
  author  = {Chis-Ciure, R.},
  title   = {Consciousness science and constitutive a priori principles: On the fundamental identity of integrated information theory},
  journal = {Philosophical Explorations},
  year    = {2025},
  volume  = {28},
  number  = {3},
  pages   = {303--327},
  doi     = {10.1080/13869795.2025.2550245},
}

@article{chisCiure2026inferential,
  author  = {Chis-Ciure, R.},
  title   = {Inferencialmente degenerado},
  journal = {Revista de Occidente},
  year    = {2026},
  volume  = {538},
  pages   = {66--79},
  issn    = {0034-8635},
}

@article{chisCiureEllia2023hardProblem,
  author  = {Chis-Ciure, R. and Ellia, F.},
  title   = {Facing up to the hard problem of consciousness as an integrated information theorist},
  journal = {Foundations of Science},
  year    = {2023},
  volume  = {28},
  number  = {1},
  pages   = {255--271},
  doi     = {10.1007/s10699-020-09724-7},
}

@article{chisCiureLevin2025,
  author  = {Chis-Ciure, R. and Levin, M.},
  title   = {Cognition all the way down 2.0: Neuroscience beyond neurons in the diverse intelligence era},
  journal = {Synthese},
  year    = {2025},
  volume  = {206},
  number  = {5},
  pages   = {257},
  doi     = {10.1007/s11229-025-05319-6},
}

@article{chisCiure2022transcendental,
  author  = {Chis-Ciure, R.},
  title   = {The transcendental deduction of integrated information theory: Connecting the axioms, postulates, and identity through categories},
  journal = {Synthese},
  year    = {2022},
  volume  = {200},
  number  = {3},
  pages   = {236},
  doi     = {10.1007/s11229-022-03704-z},
}

@phdthesis{chisCiure2022dissertation,
  author  = {Chis-Ciure, R.},
  title   = {The a priori foundations of integrated information theory: Toward a transcendental science of consciousness},
  school  = {University of Bucharest},
  address = {Bucharest},
  year    = {2022},
  type    = {Doctoral dissertation},
}

@article{elliaChisCiure2022,
  author  = {Ellia, F. and Chis-Ciure, R.},
  title   = {Consciousness and complexity: Neurobiological naturalism and integrated information theory},
  journal = {Consciousness and Cognition},
  year    = {2022},
  volume  = {100},
  pages   = {103281},
  doi     = {10.1016/j.concog.2022.103281},
}

@article{zheng2026loops,
  author  = {Zheng, Z. and Chis-Ciure, R. and Waade, P. T. and Eiserbeck, A. and Aru, J. and Andrillon, T. and Jarraya, B. and Melloni, L. and Northoff, G. and Rosas, F. E. and Dwarakanath, A.},
  title   = {Loops all the way up: Recurrency as an implementation-first primitive for consciousness},
  journal = {Physics of Life Reviews},
  year    = {2026},
  volume  = {58},
  pages   = {360--381},
  doi     = {10.1016/j.plrev.2026.07.006},
}

@article{cogitate2025,
  author  = {{Cogitate Consortium} and Ferrante, O. and Gorska-Klimowska, U. and Henin, S. and Hirschhorn, R. and Khalaf, A. and Lepauvre, A. and Liu, L. and Richter, D. and Vidal, Y. and Bonacchi, N. and Brown, T. and Sripad, P. and Armendariz, M. and Bendtz, K. and Ghafari, T. and Hetenyi, D. and Jeschke, J. and Kozma, C. and others},
  title   = {Adversarial testing of global neuronal workspace and integrated information theories of consciousness},
  journal = {Nature},
  year    = {2025},
  volume  = {642},
  pages   = {133--142},
  doi     = {10.1038/s41586-025-08888-1},
}

@article{corcoran2023,
  author  = {Corcoran, A. W. and Hohwy, J. and Friston, K. J.},
  title   = {Accelerating scientific progress through {Bayesian} adversarial collaboration},
  journal = {Neuron},
  year    = {2023},
  volume  = {111},
  pages   = {3505--3516},
  doi     = {10.1016/j.neuron.2023.08.027},
}

@article{cronbachMeehl1955,
  author  = {Cronbach, L. J. and Meehl, P. E.},
  title   = {Construct validity in psychological tests},
  journal = {Psychological Bulletin},
  year    = {1955},
  volume  = {52},
  number  = {4},
  pages   = {281--302},
}

@article{fazekas2024,
  author  = {Fazekas, P. and Cleeremans, A. and Overgaard, M.},
  title   = {A construct-first approach to consciousness science},
  journal = {Neuroscience and Biobehavioral Reviews},
  year    = {2024},
  volume  = {156},
  pages   = {105480},
  doi     = {10.1016/j.neubiorev.2023.105480},
}

@article{grim2022,
  author  = {Grim, P. and Seidl, F. and McNamara, C. and Rago, H. E. and Astor, I. N. and Diaso, C. and Ryner, P.},
  title   = {Scientific theories as {Bayesian} nets: Structure and evidence sensitivity},
  journal = {Philosophy of Science},
  year    = {2022},
  volume  = {89},
  pages   = {42--69},
  doi     = {10.1017/psa.2021.18},
}

@article{hand1996,
  author  = {Hand, D. J.},
  title   = {Statistics and the theory of measurement},
  journal = {Journal of the Royal Statistical Society: Series A (Statistics in Society)},
  year    = {1996},
  volume  = {159},
  number  = {3},
  pages   = {445--492},
  doi     = {10.2307/2983326},
}

@article{hemming2018,
  author  = {Hemming, V. and Walshe, T. V. and Hanea, A. M. and Fidler, F. and Burgman, M. A.},
  title   = {Eliciting improved quantitative judgements using the {IDEA} protocol: A case study in natural resource management},
  journal = {PLOS ONE},
  year    = {2018},
  volume  = {13},
  number  = {6},
  pages   = {e0198468},
  doi     = {10.1371/journal.pone.0198468},
}

@article{henderson2010,
  author  = {Henderson, L. and Goodman, N. D. and Tenenbaum, J. B. and Woodward, J. F.},
  title   = {The structure and dynamics of scientific theories: A hierarchical {Bayesian} perspective},
  journal = {Philosophy of Science},
  year    = {2010},
  volume  = {77},
  number  = {2},
  pages   = {172--200},
  doi     = {10.1086/651319},
}

@article{iitConcerned2025,
  author  = {{IIT-Concerned} and Klincewicz, M. and Cheng, T. and Schmitz, M. and Sebasti{\'a}n, M. {\'A}. and Snyder, J. S.},
  title   = {What makes a theory of consciousness unscientific?},
  journal = {Nature Neuroscience},
  year    = {2025},
  note    = {Advance online publication},
  doi     = {10.1038/s41593-025-01881-x},
}

@article{kirkebyHinrup2024,
  author  = {Kirkeby-Hinrup, A.},
  title   = {Quantifying empirical support for theories of consciousness: A tentative methodological framework},
  journal = {Frontiers in Psychology},
  year    = {2024},
  volume  = {15},
  pages   = {1341430},
  doi     = {10.3389/fpsyg.2024.1341430},
}

@book{lakatos1978,
  author    = {Lakatos, I.},
  title     = {The Methodology of Scientific Research Programmes: Philosophical Papers, Volume 1},
  publisher = {Cambridge University Press},
  year      = {1978},
  note      = {Edited by J. Worrall and G. Currie},
}

@article{melloni2023,
  author  = {Melloni, L. and Mudrik, L. and Pitts, M. and Bendtz, K. and Ferrante, O. and Gorska, U. and Hirschhorn, R. and Khalaf, A. and Kozma, C. and Lepauvre, A. and Liu, L. and Mazumder, D. and Richter, D. and Zhou, H. and Blumenfeld, H. and Boly, M. and Chalmers, D. J. and Devore, S. and Fallon, F. and others},
  title   = {An adversarial collaboration protocol for testing contrasting predictions of global neuronal workspace and integrated information theory},
  journal = {PLOS ONE},
  year    = {2023},
  volume  = {18},
  number  = {2},
  pages   = {e0268577},
  doi     = {10.1371/journal.pone.0268577},
}

@article{myungPitt2009,
  author  = {Myung, J. I. and Pitt, M. A.},
  title   = {Optimal experimental design for model discrimination},
  journal = {Psychological Review},
  year    = {2009},
  volume  = {116},
  number  = {3},
  pages   = {499--518},
  doi     = {10.1037/a0016104},
}

@article{negro2024,
  author  = {Negro, N.},
  title   = {(Dis)confirming theories of consciousness and their predictions: Towards a {Lakatosian} consciousness science},
  journal = {Neuroscience of Consciousness},
  year    = {2024},
  volume  = {2024},
  number  = {1},
  pages   = {niae012},
  doi     = {10.1093/nc/niae012},
}

@misc{negroEtAl2026,
  author = {Negro, N. and Elbaz, E. and Fischer, S. and Schreiber, M. and Mudrik, L.},
  title  = {Prediction maps as a gateway to better understand and test scientific theories},
  year   = {2026},
  note   = {PsyArXiv preprint},
  doi    = {10.31234/osf.io/kqr87\_v1},
}

@article{corcoran-haun-et-al-2026,
    title = {Integrated information and predictive processing theories of consciousness: {An} adversarial collaborative review},
    volume = {187},
    issn = {0149-7634},
    shorttitle = {Integrated information and predictive processing theories of consciousness},
    doi = {10.1016/j.neubiorev.2026.106742},
    urldate = {2026-08-30},
    journal = {Neuroscience \& Biobehavioral Reviews},
    author = {Corcoran, Andrew W. and Haun, Andrew M. and Dorman, Reinder and Tononi, Giulio and Friston, Karl J. and Pennartz, Cyriel M. A.},
    month = aug,
    year = {2026},
    pages = {106742},
}

@article{oizumi2014,
  author  = {Oizumi, M. and Albantakis, L. and Tononi, G.},
  title   = {From the phenomenology to the mechanisms of consciousness: {Integrated Information Theory} 3.0},
  journal = {PLoS Computational Biology},
  year    = {2014},
  volume  = {10},
  number  = {5},
  pages   = {e1003588},
  doi     = {10.1371/journal.pcbi.1003588},
}

@article{albantakis2023,
  author  = {Albantakis, L. and Barbosa, L. and Findlay, G. and Grasso, M. and Haun, A. M. and Marshall, W. and Mayner, W. G. P. and Zaeemzadeh, A. and Boly, M. and Juel, B. E. and Sasai, S. and Fujii, K. and David, I. and Hendren, J. and Lang, J. P. and Tononi, G.},
  title   = {Integrated information theory ({IIT}) 4.0: Formulating the properties of phenomenal existence in physical terms},
  journal = {PLOS Computational Biology},
  year    = {2023},
  volume  = {19},
  number  = {10},
  pages   = {e1011465},
  doi     = {10.1371/journal.pcbi.1011465},
}

@article{kleiner2020,
  author  = {Kleiner, J.},
  title   = {Mathematical models of consciousness},
  journal = {Entropy},
  year    = {2020},
  volume  = {22},
  number  = {6},
  pages   = {609},
  doi     = {10.3390/e22060609},
}

@article{boly2017,
  author  = {Boly, M. and Massimini, M. and Tsuchiya, N. and Postle, B. R. and Koch, C. and Tononi, G.},
  title   = {Are the neural correlates of consciousness in the front or in the back of the cerebral cortex? {Clinical} and neuroimaging evidence},
  journal = {The Journal of Neuroscience},
  year    = {2017},
  volume  = {37},
  number  = {40},
  pages   = {9603--9613},
  doi     = {10.1523/JNEUROSCI.3218-16.2017},
}

@article{odegaard2017,
  author  = {Odegaard, B. and Knight, R. T. and Lau, H.},
  title   = {Should a few null findings falsify prefrontal theories of conscious perception?},
  journal = {The Journal of Neuroscience},
  year    = {2017},
  volume  = {37},
  number  = {40},
  pages   = {9593--9602},
  doi     = {10.1523/JNEUROSCI.3217-16.2017},
}
\endgroup

\end{document}